\documentstyle[aps,preprint]{revtex}

\newcommand{\be}{\begin{equation}}
\newcommand{\ee}{\end{equation}}
\newcommand{\ba}{\begin{eqnarray}}
\newcommand{\ea}{\end{eqnarray}}

\begin{document}

\begin{center}
\begin{tabular}{c}
    {\Large \bf Management of the Angular Momentum of Light:  Preparation of Photons in Multidimensional Vector States of Angular Momentum
}\\
\end{tabular}
\end{center}
\vspace{1.5cm}

\begin{center}
{\bf Gabriel Molina-Terriza, Juan P. Torres, and Lluis Torner}\\
Laboratory of Photonics, Department of Signal Theory and Communications,\\
Universitat Politecnica de Catalunya, Gran Capitan UPC-D3, Barcelona, ES~08034,
Spain
\end{center}
\vspace{.5cm}

\noindent
\begin{abstract}
We put forward schemes to prepare photons in multi-dimensional vector states of
orbital angular momentum. We show realizable light distributions that yield
prescribed states with finite or infinite normal modes. In particular, we show
that suitable light vortex-pancakes allow the add-drop of specific vector
projections. We suggest that such photons might allow the generation of
{\it engineered quNits\/} in  multi-dimensional quantum information systems.
\end{abstract}
\vspace{2cm}

\vspace{2cm}

\noindent
{\bf Phys Rev Lett }

\newpage

Light carries energy and both, linear and angular momenta. The total angular
momentum can contain a spin contribution associated with polarization
\cite{Poy09,Bet36}, and an orbital contribution associated with the spatial
profile of the light intensity and phase \cite{AllPadBab99}. Such angular
momentum can be transferred to trapped suitable material particles causing them
to rotate, a property with important applications in optical tweezers and
spanners in fields as diverse as biosciences \cite{Ash99} and micromechanics
\cite{GalOrm01}. The angular momentum of light can also be used to encode
quantum information that is carried by the corresponding photon states
\cite{MaiVazWei01}.  In 
this regard, exploitation of the orbital contribution to the angular momentum
opens the door to the generation and manipulation of multi-dimensional quantum
entangled states \cite{RecZeiBer94}, with an
arbitrarily large number of entanglement dimensions. Such multi-dimensional
entanglement should allow the exploration of deeper quantum features, and might
guide the elucidation of capacity-increased quantum information processing
schemes.

Allen and co-workers showed a decade ago that paraxial Laguerre-Gaussian laser
beams carry a well-defined orbital angular momentum associated to their spiral
wave fronts \cite{AllBeiSpr92}. The formal analogy between paraxial optics and
quantum mechanics implies that such modes are the eigenmodes of the quantum
mechanical angular momentum operator \cite{CohDiuLal77,EnkNie92}.  The
Laguerre-Gaussian modes form a complete Hilbert set and can thus be used to
represent the quantum photon states within the paraxial regime of light
propagation. The quantum angular momentum number carried by the photon is then
represented by the topological charge, or winding number $m$, of the
corresponding mode, and each mode carries an orbital angular momentum of
$m\hbar$ per photon. In this Letter, we show how to construct multi-dimensional
vector photon states with controllable projections into modes with well-defined
winding numbers. In particular, we put forward schemes based on suitable light
vortex-pancakes made of Gaussian laser beams with distributions of
nested topological screw wave front dislocations, which allow the manipulation,
including the addition and removal, of specific projections of the vector
states.  We also anticipate that the schemes developed by Nienhuis, Courtial
and co-workers \cite{Nie96} to measure the rotational frequency shift
imparted to a vortex light beam, might be employed to verify our predictions.

Only the total angular momentum, containing spin and orbital contributions, is
a quantum mechanical physical observable.  Here we restrict ourselves to the
orbital contribution.  We thus consider the slowly varying electric field
envelope $u(x,y;z)$ of a cw paraxial beam propagating in free-space; $z$ is the
propagation direction, $x$ and $y$ are the spatial transverse coordinates. The
time averaged energy per unit length carried by the beam is $U=2 \epsilon_0
 \int \int |u|^2\, \mbox{d}x\mbox{d}y$, where $\epsilon_0$ is the
permitivity of vacuum. The time averaged
$z$-component of the orbital angular momentum per unit length carried by the
light beam is given by $L_z=\int \int \left[ \vec{r}_{\perp} \times
\vec{p}\right] \,
\mbox{d}x\mbox{d}y$, where $\vec{r}_{\perp}$ is the vector position in the X-Y
plane, $\vec{p}=(i \epsilon_0/\omega) \left[ u \nabla_{\perp} u^{*}-u^{*}
\nabla_{\perp} u \right]$, and $\omega$ is the angular frequency.  The
Laguerre-Gauss (LG) modes $u_{mp}$ form a complete, infinite-dimensional basis
for the solutions of the paraxial wave equation, thus any field distribution
can be represented as a vector state in that basis.  They are characterized by
the two integer indices $p$ and $m$. The index $p$ can take any non-negative
value and determines the radial shape, or node number, of the light
distribution. The index $m$ can take any integer number and determines the
azimuthal phase dependence of the mode.  When $m \ne 0$ the LG modes contain
screw wave front dislocations, or optical vortices, with topological charge $m$
nested on them.  To elucidate the angular momentum content of a field
distribution $u(x,y;z)$ one has to compute its projection into the spiral
harmonics $\exp(i n \varphi)$
\cite{CohDiuLal77}, where $n$ is the winding number. We thus let
\begin{equation}
u(\rho,\varphi;z)=\frac{1}{\sqrt{2 \pi}}\,
        \sum_{n=-\infty}^{\infty} a_n(\rho,z) \,\exp (i n \varphi),
\end{equation}
where $a_n(\rho,z)=1/(2 \pi)^{1/2} \int_{0}^{2 \pi}
u(\rho,\varphi,z) \exp(-i n \varphi)\,\mbox{d}\varphi$. 
The energy carried by the corresponding light beam
can be written as $U=2 \epsilon_0 \sum_{-\infty}^{\infty} C_n$,
 where  $C_n=\int_{0}^{\infty} |a_n(\rho,z)|^2\,\rho \, \mbox{d}\rho$, can be
shown to be constants independent of $z$.  The angular momentum carried by the
light beam is thus given by $L_z=(2\epsilon_0/\omega)\,\sum_{-\infty}^{\infty}
n\, C_n.$ When the energy $U$ is measured in units of $\hbar
\omega$ (i.e., $\tilde{U}=U/\hbar
\omega$), the ratio  $\tilde{L}_z=L_z/\tilde{U}$, which is usually referred as
the angular momentum 
per photon \cite{AllPadBab99}, is given by $\hbar\,{\sum_{-\infty}^{\infty}n
C_n}/{\sum_{-\infty}^{\infty} C_n}$. 

Within the paraxial regime of light propagation the LG modes are the eigenmodes
of the quantum mechanical energy $\hat{E}$ and angular momentum $\hat{L}_z$
operators \cite{AllPadBab99,CohDiuLal77,EnkNie92}, so photons represented by a
single LG mode are in a quantum state $|m>$ with well defined values of energy
($\hat{E}\,|m>\,=\,\hbar \omega\,|m>$) and orbital angular momentum
($\hat{L}_z\,|m>\,=\,m \hbar\,|m>$). State vectors which are not represented by
a pure LG mode correspond to photons in a superposition state, and the weights
of the quantum superposition $\{P_n\}$ are determined by the array $\{C_n\}$,
by the expression
\begin{equation}
\label{pn}
P_n=\frac{C_n}{\sum_{-\infty}^{\infty} C_l}.
\end{equation}
The mean value of the orbital angular momentum per photon, obtained by a full
quantum average over many realizations, is
$\tilde{L}_z=\hbar\,\sum_{-\infty}^{\infty} n P_n$.  The central
idea put forward in this paper is to control the series $\{P_n\}$, hence to
prepare photons in a superposition state of modes $|m>$, by elucidating
suitable light field distributions.  Such superposition can be
restricted to a finite number of modes, which are to be chosen, or it can
consist of an infinite, but discrete, number of modes.

Photons that carry angular momentum in a superposition state of an infinite,
but controllable, number of normal modes can be prepared in a variety of ways.
An experimentally important scheme is obtained by passing a pure state $|m>$
through astigmatic optical components \cite{EnkNie92,AbrVol91}.  A
light beam prepared in a pure $m$-order LG mode with the beam center located at
$(x,y)=(0,0)$, but whose orbital angular momentum is measured relative to an
origin located at $(x_0, 0)$, constitutes another important example of an
infinite-dimensional superposition state.  The angular momentum operator
$\hat{L_z}$ relative to the displaced origin is given by $\hat{L_z}=
\hat{L}_{oz}+ \exp(i x_0 \hat{P}_x) \left[ \hat{L}_{oz}, \exp(-i x_0 \hat{P}_x)
\right]$, where $\hat{L}_{oz}$ is the operator associated to the point $(0,0)$,
$\hat{P}_x$ is the linear momentum operator, and $[,]$ is a Poisson bracket.
Thus, even though the photons are prepared in a pure $|m>$ state for the
operator $\hat{L}_{oz}$, such is not the case for the operator $\hat{L}_{z}$.
A direct extension of the above are multi-pearl necklace light fields
\cite{SolSeg00}, but wih nested vortices in each pearl \cite{MolRecTor00}. Consider the simplest
case of a two-pearl vortex-necklace with $u(x,y,z)=A_{m0}
u_{m0}(x+x_0/2,y,z)+B_{m0} u_{m0}(x-x_0/2,y,z)$, which corresponds to the
coherent superposition of two LG beams separated a distance $d=x_0$. The
relative amplitudes and phases between both modes, and the length $d$, are the
control parameters to act on $\{P_n\}$.  Let $m=1$, and $A_{10}=B_{10}$. Then, when
the pearls are superimposed ($d=0$), photons are prepared in a pure $|m=1>$
state.  However, when $d\ne 0$ photons are prepared in an infinite-dimensional
superposition state of {\it odd modes only}. Figure 1 shows typical examples.
Vortex-necklaces that prepare photons in states with {\it even modes only} can
also be constructed.

The ultimate goal in the effort to prepare the state of photons carrying
orbital angular momentum is to elucidate a light signal that yields a
superposition into a {\it finite, in principle arbitrary large, number of $|m>$
states}, which in addition allows {\it adding-dropping specific projections}.
In what follows, we show that such goal can be achieved by using light
vortex-pancakes made of properly distributed single-charge screw dislocations
nested into a Gaussian host. A pancake with $N$ single-charged dislocations is
given by \cite{Ind93}
\begin{equation}
u_N(\rho,\varphi;z=0)=A_0 \prod_{l=1}^{N} \left[ \rho 
        \exp (i \varphi)-\rho_l
        \exp (i \varphi_l) \right]
        \exp(-\rho^2/w_0^2),
\label{pancake}
\end{equation}
where $A_0$ is a constant, $\rho_l$, $\varphi_l$ are the radial and
azimuthal positions of the $l$-th vortex in cylindrical coordinates,
respectively, and $w_0$ is the beam waist. Projection of
(\ref{pancake}) onto LG modes $u_{mp}$ yields
\begin{equation}
\label{pancakedecomposed}
u_N(\rho,\varphi;z)=A_0 \, \sqrt{\pi}\,\sum_{l=0}^{N}
        (-1)^{N-l}
        \left( 
        \frac{w_0}{\sqrt{2}}
        \right)^{l+1}
        \sqrt{l!}\,B_{N-l}\, u_{l0}(\rho,\varphi;z),
\end{equation}
where 
$
        B_n=\sum_{j_1} \sum_{j_2}  \cdot \cdot \cdot
                \sum_{j_n} 
                \prod_{l=1}^{n} \rho_{j_l} \exp(i \varphi_{j_l}),
$
with $j_l\in [1,N]$, and $j_l<j_{l+1}$.  Computation
of the array $\{C_n\}$ for the distribution (\ref{pancake}) yields
\begin{equation}
C_n= |A_0|^2 \,\pi\, n! \,  \left( \frac{w_0^2}{2} \right)^{n+1} |B_{N-n}|^2.
\label{cn}
\end{equation}
According to Eq.~(\ref{pancakedecomposed}), the vortex-pancakes (\ref{pancake})
prepare photons in superpositions of a maximum of $N+1$, $|m>$ states.  The
actual number of states and their weights are given by the positions of the
vortices nested in the beam at $z=0$. Notice that even though such positions
vary with $z$, together with the beam waist and wave front curvature, the
weights of the $|m>$ states do not. Calculation of the weights $\{P_n\}$
associated to a given light distribution is straightforward (even though in
general it has to be done numerically), but the interesting problem is just the
opposite: Finding a light vortex-pancake that yields a desired distribution
$\{P_n\}$. In the case of small $N$, the problem can be solved in analytical
form.  $N=1$ corresponds to a Gaussian beam with a nested off-axis vortex
\cite{MaiVazWei01,VasSlySos01}, and 
prepares photons in a superposition of the states $|m=0>$ and $|m=1>$, only. One finds that the mean value of the angular momentum per
photon can take any value between $0$ and $\hbar$. The two limiting cases
correspond to the vortex located at the center of the host beam (then
$\tilde{L}_z=\hbar$), and to the vortex located very far from the center, which
obviously yields $\tilde{L}_z\rightarrow 0$. We will consider in detail the
case $N=2$ which corresponds to a Gaussian beam with waist $w_0$ with two
single-charge vortices nested off-axis. The corresponding photons are set in
3-dimensional vector states, and the elements of the array $\{C_n\}$ are found to
be given by
\begin{eqnarray}
& & C_{0}= \frac{1}{2}\,w_0^2 \pi \rho_1^2 \rho_2^2\,|A_0|^2, \\ 
&  & C_{1}= \frac{1}{4}\,w_0^4\,\pi\,|A_0|^2\, 
        \left[ \rho_1^2 + \rho_2^2 +2 \rho_1 \rho_2 \cos
     (\varphi_1-\varphi_2) \right], \\  
& & C_{2}=\frac{1}{4}\, w_0^6 \pi\,|A_0|^2.
\end{eqnarray}
Most vortex locations lead to photons states with the three possible dimensions
occupied, so that $P_0 \ne 0$, $P_1 \ne 0$, $P_2 \ne 0$. For example,
equi-distributed populations, with $P_0=P_1=P_2=1/3$, are obtained with a
pancake where the vortices are located at
\begin{eqnarray}
       & & \rho_1^2\rho_2^2=\frac{w_0^4}{2},\\
       & & \varphi_1-\varphi_2=\pi-\cos^{-1}
        \left(
        \frac{w_0^2-\rho_1^2-\rho_2^2}{2 \rho_1 \rho_2}
        \right).
\end{eqnarray}
Photon states with one of the projections suppressed can also be realized.
One gets {\bf $P_0=0$} states by letting $\rho_1=0$. Then,
any $(P_1,P_2)$ arbitrary combination, with $P_1+P_2=1$, is obtained when 
$\rho_{2}=w_0 [(1-P_2)/P_2]^{1/2}$.
{\bf $P_1 = 0$} states can be realized by letting
$\varphi_1-\varphi_2=\pi$, and $\rho_1=\rho_2$. Then, 
any $(P_0,P_2)$ combination, with $P_0+P_2=1$, is obtained when
$\rho_{1}=\rho_2 =w_0 [(1-P_2)/2P_2]^{1/4}$.
{\bf $P_2\rightarrow 0$} states can only be achieved
asymptotically, by locating one of the vortices very far from the beam center
(i.e., $\rho_2 \rightarrow \infty$). 
Then, 
any $(P_0,P_1)$ combination, with $P_0+P_1\rightarrow 1$, is obtained when the
remaining off-axis vortex is located  at
$\rho_{1}= w_0 [(1-P_1)/2P_1]^{1/2}$.
Figure 2 illustrates the actual light vortex-pancakes and the corresponding
$\{P_n\}$ weights for the above cases. In principle,  multi-dimensional
vector states with large values of $N$ can be managed by solving
numerically the inverse problem posed by (\ref{pn}), (\ref{cn}).
At present we cannot give a definite solving scheme for arbitrary $N$.
However, the plots displayed in Fig.~3, which correspond to $N=10$, show that
preparation of fairly large multi-dimensional $L$-managed photon states is
possible indeed.

To experimentally verify our predictions, a scheme that resolves the LG
spectrum of a light beam must be elucidated.  We anticipate that a set-up
similar to that implemented by Courtial and co-workers
\cite{Nie96} should serve that purpose.  Courtial et.~al.\ set-up uses a
rotating Dove 
prism that imparts a Doppler frequency shift to the light signal given by $\Delta
\omega_m=2 m \Omega$, where $\Omega$ is the angular velocity of the prism and
$m$ the topological charge of the vortex.  Thus, dislocations with different
topological charges $m$ induce different frequency shifts. Therefore, the
superposition of LG modes discussed in this paper is expected to generate a
frequency spectrum consisting of sidebands, with an amplitude given by the mode
weight, around the unshifted frequency.  Resolving such sidebands, e.g., by a
Michelson-Morley interferometer, should reveal the LG spectrum of the light
signal.

While looking for such experimental verification, we conclude noticing the
implications of the results to the generation of quantum entangled photons with
angular momentum \cite{MaiVazWei01}, by parametric down-conversion of
vortex-signals in quadratic nonlinear crystals \cite{vortex-OPA}. The idea is
to generate entangled states $\psi=\sum_{m_1, m_2} B_{m_1 m_2} |m_1>\,|m_2>$,
with modified $B_{m_1 m_2}$ probability amplitudes, by using $L$-managed pump
photons.  Such multi-dimensional entanglement should allow the experimental
exploration of quantum features only realizable in $N$-dimensional Hilbert
spaces. The degree of violation of Bell's inequalities as a function of $N$
\cite{KasGnaZuk00}, generated by different $L$-managed vector states,
constitutes a fascinating possibility. Finally, we anticipate that $L$-managed
photon states should find applications in spintronics \cite{GanIvcDan01}
and in capacity-increased quantum information schemes
\cite{BouEkeZei00,KniLafMil01}.

This  work was supported
by the Generalitat de Catalunya and by TIC2000-1010.  Numerics were carried out
at CIRI.

\newpage

\noindent
{\large\bf \underline{FIGURE CAPTIONS}}
\vspace{1cm}

\small

\noindent
{\bf Fig.~1.} Weight of the $|m>$ states as a function of the $m$-quantum
number, for a two-pearl vortex-necklace light distribution, for
different separations $d$ between the pearls. (a): $d=0$; (b): $d=w_0$; (c):
$d=2w_0$; (d): $d=6w_0$.  Each pearl features a LG$_{10}$ shape. The insets
show the wave function amplitude and a sketch of the location of the existing
screw wave front dislocations in each case.  Notice that in (a) and (b), only
one dislocation is present in the whole wave front, whereas in (c) and (d) the
wave front contains three dislocations. The net topological charge 
is always +1.
\vspace{.5cm}

\noindent
{\bf Fig.~2.} Preparation of $L$-managed 3-dimensional photon states with a
$N=2$ vortex-pancake.  Features as in Fig.~1. In (a): $P_0=P_1=P_2=1/3$; in
(b): $P_0=0$; $P_1=P_2=1/2$;  in (c): $P_1=0$; in (d): $P_2=0$.
\vspace{.5cm}

\noindent
{\bf Fig.~3.} Analogous to Fig.~2, but for 11-dimensional photon states
prepared with a $N=10$ vortex-pancake. (a): Light amplitude; (b): Mode
weight versus azimuthal location of one of the vortices (in the particular case
shown, $\varphi_1$); (c)-(d): Mode weigth versus $m$-quantum number for two
different pancakes. (c) corresponds to (a), and in (d) $\varphi_1$ was chosen
so that the $|m=4>$ mode is almost suppressed, as dictated by (b).

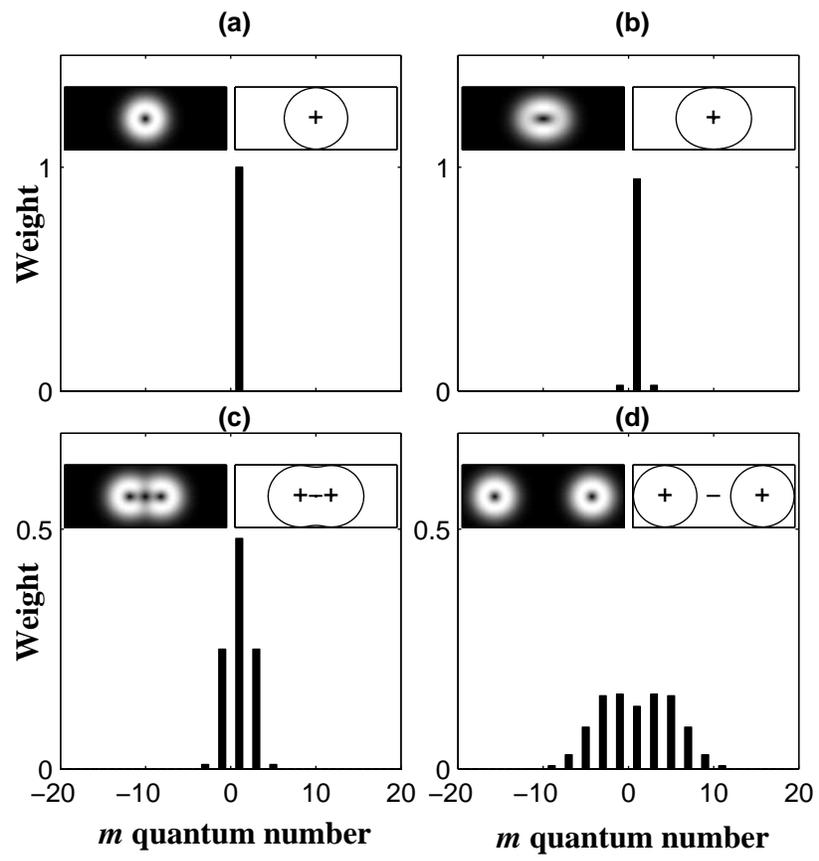
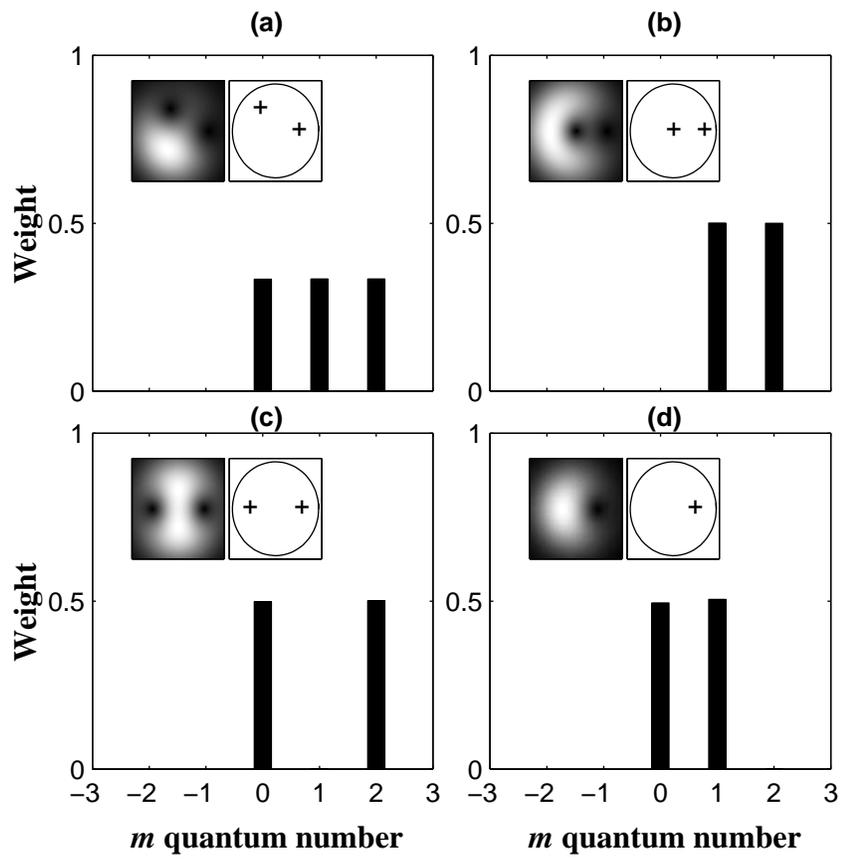
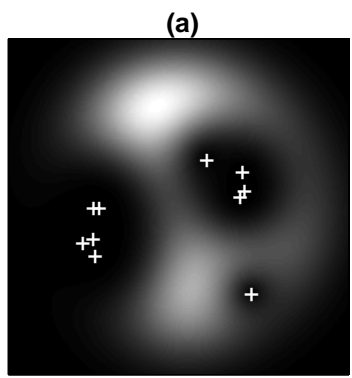
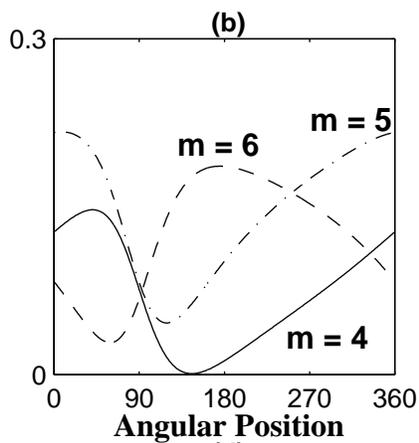
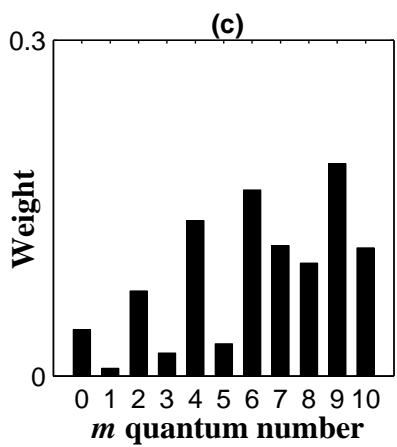
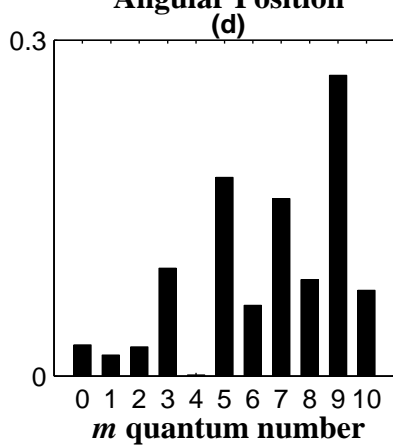
\end{document}